\documentclass[sigconf]{acmart}

\usepackage{changepage}
\usepackage{float}
\usepackage{stfloats}
\AtBeginDocument{%
  \providecommand\BibTeX{{%
    \normalfont B\kern-0.5em{\scshape i\kern-0.25em b}\kern-0.8em\TeX}}}

\copyrightyear{2024}
\acmYear{2024}
\setcopyright{acmlicensed}\acmConference[CHI '24]{Proceedings of the CHI Conference on Human Factors in Computing Systems}{May 11--16, 2024}{Honolulu, HI, USA}
\acmBooktitle{Proceedings of the CHI Conference on Human Factors in Computing Systems (CHI '24), May 11--16, 2024, Honolulu, HI, USA}
\acmDOI{10.1145/3613904.3641899}
\acmISBN{979-8-4007-0330-0/24/05}

\begin{document}

\title[ABScribe: Exploring \& Organizing Multiple Writing Variations in Human-AI Co-Writing Tasks]{ABScribe: Rapid Exploration \& Organization of Multiple Writing Variations in Human-AI Co-Writing Tasks using Large Language Models}

\author{Mohi Reza}
\email{mohireza@cs.toronto.edu}
\orcid{0000-0001-9668-3384}
\affiliation{
  \institution{University of Toronto}
  \city{Toronto}
  \state{Ontario}
  \country{Canada}
  \postcode{M5S 2E4}
}

\author{Nathan Laundry}
\email{nathan.laundry@mail.utoronto.ca}
\orcid{0000-0003-0846-2472}
\affiliation{
  \institution{University of Toronto}
  \city{Toronto}
  \state{Ontario}
  \country{Canada}
  \postcode{M5S 2E4}
}

\author{Ilya Musabirov}
\email{ilya.musabirov@mail.utoronto.ca}
\orcid{0000-0003-2246-0094}
\affiliation{
  \institution{University of Toronto}
  \city{Toronto}
  \state{Ontario}
  \country{Canada}
  \postcode{M5S 2E4}
}

\author{Peter Dushniku}
\email{peter.dushniku@mail.utoronto.ca}
\orcid{0009-0002-3789-4629}
\affiliation{
  \institution{University of Toronto}
  \city{Toronto}
  \state{Ontario}
  \country{Canada}
  \postcode{M5S 2E4}
}

\author{Michael Yu}
\email{michaelz.yu@mail.utoronto.ca}
\orcid{0009-0008-7255-9838}
\affiliation{
  \institution{University of Toronto}
  \city{Toronto}
  \state{Ontario}
  \country{Canada}
  \postcode{M5S 2E4}
}

\author{Kashish Mittal}
\email{kashish.mittal@mail.utoronto.ca}
\orcid{0009-0000-2835-3797}
\affiliation{
  \institution{University of Toronto}
  \city{Toronto}
  \state{Ontario}
  \country{Canada}
  \postcode{M5S 2E4}
}

\author{Tovi Grossman}
\email{tovi@dgp.toronto.edu}
\orcid{0000-0002-0494-5373}
\affiliation{
  \institution{University of Toronto}
  \city{Toronto}
  \state{Ontario}
  \country{Canada}
  \postcode{M5S 2E4}
}

\author{Michael Liut}
\email{michael.liut@utoronto.ca}
\orcid{0000-0003-2965-5302}
\affiliation{%
  \institution{University of Toronto Mississauga}
  \city{Mississauga}
  \state{Ontario}
  \country{Canada}
  \postcode{L5L 1C6}
}

\author{Anastasia Kuzminykh}
\email{anastasia.kuzminykh@utoronto.ca}
\orcid{0000-0002-5941-4641}
\affiliation{%
  \institution{University of Toronto}
  \city{Toronto}
  \state{Ontario}
  \country{Canada}
  \postcode{M5S 3G6}
}

\author{Joseph Jay Williams}
\email{williams@cs.toronto.edu}
\orcid{0000-0002-9122-5242}
\affiliation{%
  \institution{University of Toronto}
  \city{Toronto}
  \state{Ontario}
  \country{Canada}
  \postcode{M5S 2E4}
}

\renewcommand{\shortauthors}{Reza, et al.}

\begin{abstract}
  Exploring alternative ideas by rewriting text is integral to the writing process. State-of-the-art Large Language Models (LLMs) can simplify writing variation generation. However, current interfaces pose challenges for simultaneous consideration of multiple variations: creating new variations without overwriting text can be difficult, and pasting them sequentially can clutter documents, increasing workload and disrupting writers' flow. To tackle this, we present ABScribe, an interface that supports rapid, yet visually structured, exploration and organization of writing variations in human-AI co-writing tasks. With ABScribe, users can swiftly modify variations using LLM prompts, which are auto-converted into reusable buttons. Variations are stored adjacently within text fields for rapid in-place comparisons using mouse-over interactions on a popup toolbar. Our user study with 12 writers shows that ABScribe significantly reduces task workload ($d = 1.20, p < 0.001$), enhances user perceptions of the revision process ($d = 2.41, p < 0.001$) compared to a popular baseline workflow, and provides insights into how writers explore variations using LLMs.
\end{abstract}



\begin{teaserfigure}
\centering
\includegraphics[width=0.86\textwidth]{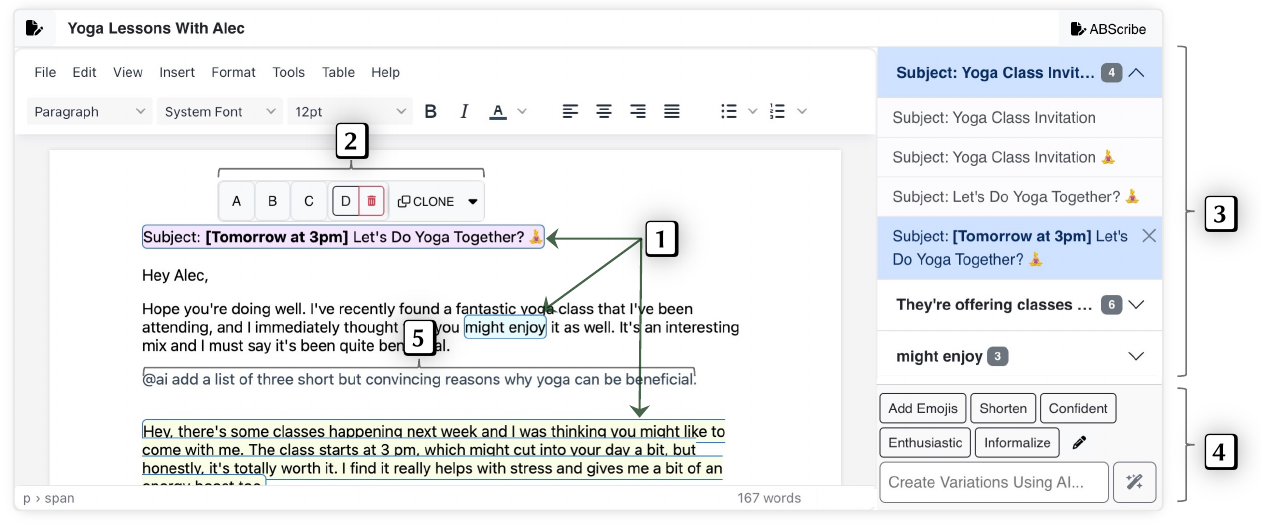}
  \caption{The ABScribe Interface: (1) \textbf{Variation Fields}: Multiple variations are stored within text fields that do not break the flow of the draft. (2) \textbf{Popup Toolbar}: Users can swiftly choose and compare multiple variations by hovering over a dynamically placed toolbar located above the selected Variation Field, or clone and edit variations in-place. (3) \textbf{Variation Sidebar:} Users can navigate through multiple variations organized as an accordion-based interface. (4) \textbf{AI Modifiers}: Users can create variations using AI by typing modification instructions auto-converted into reusable buttons that can be applied to other Variation Fields. (5) \textbf{AI Drafter}: Users can draft text from GPT-4 directly into the document by typing `@ai <prompt>' and pressing enter.}
  \Description{Components of the ABScribe Interface}
  \label{fig:teaser}
\end{teaserfigure}



\maketitle

\section{Introduction}

\begin{quote}
\centering
    \textit{``The only kind of writing is rewriting''} -- Ernest Hemingway, \textit{A Moveable Feast} \cite{hemingway2014moveable}
\end{quote}

Revision is essential to the writing process \cite{zinsser2006writing, hemingway2014moveable, goldberg2016writing, mcphee2017draft}. Professional writers often write and rewrite text hundreds of times \cite{sommers1980revision, chenoweth1987need} and  recommend rewriting as a core strategy for writing well \cite{zinsser2006writing, chenoweth1987need}. Effective revision goes beyond minor editorial changes, and may help writers rework ideas, and powerfully affect their knowledge \cite{fitzgerald1987research, sommers1980revision} as they explore alternative variations to find a line of argument \cite{fitzgerald1987research}. The revision process is \textit{iterative} \cite{du-etal-2022-understanding-iterative, fitzgerald1987research}: happens in repeated cycles, throughout the writing process, \textit{granular} \cite{seow2002writing, mcphee2017draft}: happens at the word, sentence, or paragraph-level, and \textit{non-linear}\cite{cheung2016teaching, sommers1980revision}: requires constant reconsideration of potential variations of existing text throughout the passage.

Current writing interfaces tend not to support a non-linear revision process and predominantly support linear representations of revision history (e.g., revision history in popular word processing software such as Google docs or Microsoft Word). While these tools do support iterative and granular edits, it remains difficult for writers to simultaneously consider multiple writing variations and to organize them without replacing earlier text, or cluttering documents when writers resort to pasting them in sequence. Insights from HCI and traditional design practice suggest that simultaneous consideration of multiple (at least 5) variations can lead to better ideas \cite{10.1145/1124772.1124960} and avoid fixation \cite{jansson1991design, 10.1145/1124772.1124960, 10.1145/1357054.1357074, buxton2010sketching}. We hypothesize that this may apply to writing, provided that writers are given adequate support in managing multiple variations with minimal workload. 

Advanced Large Language Models (LLMs) such as ChatGPT \cite{chatgpt}, GPT-4 \cite{gpt-4}, PaLM 2 \cite{PaLM2}, LLaMA 2 \cite{touvron2023llama}, and Gemini{\cite{Pichai_Hassabis_2023}} can enable writers to generate multiple variations of text via natural language prompting \cite{brown2020language, dale2021gpt, gero2022sparks, min2023recent}, potentially reducing the workload of generating text variations. However, easier generation can exacerbate challenges surrounding the systematic organization, comparison, and modification of multiple variations with existing chat-based and in-place editing interfaces where users are required to find text variations in linear chat histories and store them in their document as comments or separate in-line text blocks. As text variations become easier to generate using AI, they become harder to manage.

Recent HCI studies on LLM-based tool design have mainly focused on prompt engineering \cite{brade2023promptify, 10.1145/3544548.3581388, liu2022design, jiang2022promptmaker, wu2022promptchainer, dang2023choice} and exploring the generative capabilities of LLMs \cite{10.1145/3490099.3511105, lee2022coauthor, 10.1145/3545947.3569630, meyer2022we, mirowski2023co}. For example, Zamfirescu-Pereira et al. investigated ways to support non-AI experts with crafting effective prompts, and Dang et al. {\cite{dang2023choice}} found that users preferred choosing from multiple suggestions and using content from the draft to guide LLM usage rather than writing external prompts which require more effort. Yuan et al,  \cite{10.1145/3490099.3511105} explored how users might use LLMs for creative writing and found that the output of the model did not need to be perfect to be useful to users. Many users found the output useful, even if they had to significantly revise the text or chose not to incorporate it into their final draft \cite{10.1145/3490099.3511105}. This highlights the potential value of designing affordances that help manage \textit{imperfect} AI-generated variations. Even if these variations don't make it into the final text, they might still be valuable to consider.

In this paper, we present ABScribe\footnote{We name our system \textit{ABScribe} to reference how we label variations using the alphabet. Note that the interface supports variations beyond just A and B. The Popup Toolbar can include multiple variations: A, B, C, D, E, etc.}–a novel writing interface that supports the rapid exploration and organization of multiple writing variations in LLM-based human-AI co-writing tasks. We draw inspiration from a design framework by Kim et al. which shows the potential for object-oriented interactions with LLMs to encourage iteration and experimentation during writing \cite{kim2023cells}, and propose an ensemble of five interface elements that support writers in swiftly working with multiple writing variations: (i) \textbf{Variation Fields} that store multiple human and AI-generated variations within flexible text segments in a non-linear manner, without overwriting text; the (ii) \textbf{Popup Toolbar} that reveal corresponding variations inside a Variation Field when users hover their mouse over buttons representing each variation, allowing for rapid comparisons without breaking text flow; (iii) the \textbf{Variation Sidebar} that organizes all variations in a navigable format within an accordion UI; (iv) \textbf{AI Modifiers} that automatically encapsulates LLM instructions into reusable buttons that can be applied across different Variation Fields; and (v) the \textbf{AI Drafter} which allows writers to insert LLM-generated text directly into the document (see Figure \ref{fig:teaser}).

To validate our design, we conducted a controlled evaluation study involving revision tasks and interviews with 12 writers comparing ABScribe with a widely-used baseline worklow consisting of an AI integrated rich text editor based on GPT-4, with a chat-based conversational AI assistant. Our findings demonstrate that ABScribe significantly reduces subjective task workload ($d = 1.20, p < 0.001, $), and enhances user perceptions of the revision process ($d = 2.\text{41}, p < 0.001$), compared to the baseline. The key contributions of our work are as follows:
\begin{enumerate}
    \item The design and implementation of ABScribe, an LLM-enhanced writing interface that supports the rapid exploration and organization of multiple text variations in human-AI co-writing tasks. We have made the full code for ABScribe open source and accessible via  GitHub\footnote{Link to GiHub Repository: \href{https://github.com/mohireza/abscribe/}{https://github.com/mohireza/abscribe/}}.
    \item The results of a 12-participant user study with writers demonstrating the efficacy of the ABScribe interface ensemble and its advantages over a commonly used baseline workflow, and user perspectives on how writers explore multiple variations in human-AI co-writing tasks using a linear and non-linear revision process.  
\end{enumerate}

\section{Related Work}

We first review literature in HCI and traditional design practices on considering multiple alternatives, and discuss how this design approach is pertinent when revising text, informed by the theory on the revision process in writing (section \ref{rel_exploring_variations}). Then, we examine this work's alignment with the rapidly growing research in AI-assisted writing, focusing on the effects of exposing writers to diverse AI-generated suggestions, and explore some challenges that arise when trying to support multiple variation editing using existing interfaces (section \ref{llm_interfaces}). Finally, we contrast between \textit{chat-based} and \textit{in-place} interfaces to situate our design within a broader class of Human-AI writing interfaces (section \ref{human-ai-interface-types}).  

\subsection{Exploring Multiple Variations}
\label{rel_exploring_variations}

HCI and traditional design practices encourage the parallel exploration of multiple variations to help avoid fixation on a singular idea \cite{jansson1991design, goldschmidt2011avoiding}, to reduce the chances of eliminating rough but innovative ideas due to premature evaluation \cite{10.1145/1357054.1357074, buxton2010sketching}, and to make us less prone to inflated subjective appraisals by giving us an opportunity to critically assess ideas in relation to each other \cite{10.1145/1124772.1124960, buxton2010sketching}. In this paper, we hypothesize that such parallel exploration of multiple variations may apply to the revision process during writing. Much like how a naive, linear implementation of an iterative design approach encourages the sequential refinement of ideas, when writers do not have a way to organize and work with multiple text variations, they may end up committing to ideas too early, and focusing too much on surface level edits to refine their draft.

This is problematic because when we turn to research on the revision and writing process, we see that revision goes beyond surface level edits \cite{fitzgerald1987research}, encompassing deeper writing subprocesses such as revising and evaluating ideas \cite{flower1981cognitive} and meaning discovery \cite{sommers1980revision}. Experienced writers treat revision as a recursive, non-linear process \cite{sommers1980revision, seow2002writing}, and engage with the text in repeated cycles, with multiple objectives including finding the form or shape of an argument \cite{sommers1980revision}, experimenting with vocabulary and style \cite{huff1983teaching}, and going back and forth between multiple composing activities during revision \cite{faigley1981analyzing}. 

In addition to the rich-body of work underscoring the important and complex role of revision in writing, researchers have explored the benefits of adopting design language in writing pedagogy, such as characterizing writing pedagogy as a wicked \cite{rittel1967wicked} design thinking problem \cite{leverenz2014design, purdy2014can}. There has also been some valuable work in HCI in designing novel editing practices, such as supporting constraints and consistency in maintaining domain-specific terms across complex documents \cite{han2020textlets} using persistent, reified \cite{10.1145/345513.345267} text selections that allow writers to store and compare two variations.  However, further innovation in this space is needed to design affordances for writers that support the simultaneous consideration of multiple variations during the revision process. 

We contribute to this line of work, and present a suite of interface elements that work together in supporting writers with the rapid exploration and organization of multiple text variations in a non-linear fashion, in alignment with the nature of the revision process, and offer empirical insights into the applicability of design ideas in HCI on the parallel consideration of variations to the specific task of revision in writing.

\subsection{Working with Multiple Variations from Large Language Models}
\label{llm_interfaces}

There has been an explosion of recent work \cite{gero2022sparks, giray2023prompt, dang2023choice, mirowski2023co, buschek2021impact, gero2023social, long2023tweetorial, chung2022talebrush, singh2023hide, kim2023towards, kim2023cells, lee2022coauthor, 10.1145/3490099.3511105, 10.1145/3532106.3533506, shakeri2021saga, goodman2022lampost} on interfaces for  writing with LLMs in tandem with rapid advancements in AI writing capabilities using LLMs \cite{min2023recent, vaswani2017attention, radford2018improving, devlin2018bert, brown2020language}, examining various aspects of the AI-assisted writing process including  prompting \cite{wu2022promptchainer, giray2023prompt}, text suggestions \cite{dang2023choice, buschek2021impact}, creative ideation \cite{gero2022sparks, chung2022talebrush}, and social dynamics \cite{10.1145/3532106.3533506, gero2023social}. Among these, prior studies on the impact of showing multiple text suggestions from the AI \cite{buschek2021impact, dang2023choice} are of particular relevance to our design. Buschek et al. \cite{buschek2021impact} highlighted the benefits of multiple parallel suggestions for ideation but warned against the simultaneous cost to efficiency. Dang et al. \cite{dang2023choice} coined and distinguished between \textit{diegetic} prompts that are part of draft with \textit{non-diegetic} prompts that are external instructions to the LLM. They uncovered user behavioural patterns indicating a tendency to choose from multiple suggestions rather than writing non-diegetic prompts as that requires more effort, and found that users prefer to use the draft, i.e., diegetic informtion, to guide LLM prompts.


AI Modifiers, for instance, simplify prompt-writing by turning modification instructions into reusable buttons, working closely with Variation Fields to offer users multiple options and utilize diegetic information directly from the draft. We also incorporate Kim et al.'s concepts of object-oriented interactions with LLMs\cite{10.1145/3313831.3376804} and draw from existing revision control research \cite{han2020textlets, han2020designing, head2019managing} to create interface elements that effectively manage multiple LLM variations, reducing task workload and facilitating parallel exploration.


\subsection{Chat-Based and In-Place Human-AI Co-Writing Interfaces}
\label{human-ai-interface-types}

To ground our interface design, we distinguish between two types of Human-AI Co-Writing interfaces: Chat-Based interfaces such as ChatGPT and In-Place interfaces that directly insert or modify text in a document. 

\textit{Chat-Based Interfaces:} Currently the dominant mode, chat-based interfaces, like ChatGPT \cite{chatgpt}, Bing Chat \cite{bing}, and Bard \cite{bard}, have gained immense popularity. These interfaces are highly intuitive, and mimic conversational interactions between humans, but lack scaffolding for crafting prompts, which can be difficult for novice AI users \cite{10.1145/3544548.3581388, kroll1994guidelines}. In the context of writing with AI, simple turn-taking between the writer and the AI does not provide adequate support to writers in steering the iterative output from the LLM \cite{chung2022talebrush}.

Another significant limitation is the linear chat-log structure. In contrast to the non-linear nature of how revision happens in writing \cite{sommers1980revision,fitzgerald1987research}, the text-variations generated using a chat-based interface are buried within linear chat-logs, impeding parallel exploration of multiple variations in-place, where the writer is editing the document. 

\textit{In-Place Editing Interfaces:} This type offers closer integration between the human and AI writer during the text editing process by adopting a more \textit{What You See Is What You Get} (WYSIWYG) \cite{bly2007fundamentals} approach where AI-generated text modifies the human text and vice versa. This offers increased flexibility over the edited content compared to chat-based interfaces by allowing users to edit individual sections of the text. 


Recent research prototypes like Wordcraft \cite{10.1145/3490099.3511105} and CoAuthor \cite{lee2022coauthor}, as well as commercial tools such as Grammarly\footnote{\href{www.grammarly.com}{grammarly.com}}, Notion AI, and Wordtune\footnote{\href{www.wordtune.com}{wordtune.com}}, offer in-place editing features for LLM writing assistance. Wordcraft \cite{10.1145/3490099.3511105} combines an editing interface with options for narrative continuation and AI-modified text replacement, while CoAuthor \cite{lee2022coauthor} leverages GPT-3 for deep interaction and multiple edit suggestions. However, these tools typically overwrite previous text with new edits, obscuring the original content. Saved versions are often linear, as seen in Google Docs, or limited to undo/redo histories, complicating the management of multiple text variations simultaneously.

In our design, we adopt an in-place editing interface in a GPT-4 powered research prototype, offering a solution to overcome challenges surrounding the management of multiple text variations in human-AI co-writing tasks. We provide fresh empirical insights from interviews with writers on how our in-place interface design compares with existing chat-based writing workflows, and their impact on user perceptions of the revision process and task workload.
\section{Designing ABScribe}
In this section, we describe the design requirements for ABScribe and the interface elements that we developed to address those requirements. 
\subsection{Design Requirements}
We surveyed literature on several key areas critical to our goal of facilitating the swift exploration of multiple writing variations using LLMs: the role and nuances of revision within the writing process \cite{sommers1980revision,fitzgerald1987research,macarthur2018evaluation,harris2017rewriting,sharples2002we,f508427a-e4c0-3d6a-8abf-03a5d21ec6c4}; HCI design philosophies that emphasize the consideration of multiple ideas before evaluation \cite{10.1145/1124772.1124960,10.1145/1357054.1357074,dow2010parallel}; principles on reification and reuse for designing visual interfaces \cite{10.1145/345513.345267,10.1145/3313831.3376804}; and the latest research on utilizing LLMs in writing interfaces to foster experimentation and creativity \cite{10.1145/3490099.3511105,kim2023cells,lee2022coauthor, buschek2021impact, dang2023choice, gero2022sparks, chung2022talebrush}. Based on this, we formulated an initial set of design requirements, which are summarized below.

\textbf{Requirement 1: Minimize task workload when exploring multiple text variations.}
Drawing from HCI and traditional design principles, we emphasize the importance of exploring multiple ideas concurrently. Instead of refining a single solution to “get the design right”, these principles encourage the iteration and evaluation of multiple solutions in parallel to ultimately “get the right design” \cite{10.1145/1124772.1124960}. We hypothesized that parallel exploration of text segments could aid writers during the revision process, in alignment with recent studies on the benefit of showing multiple parallel phrase suggestions on ideation \cite{buschek2021impact}. We also considered that during the writing process writers frequently struggle with cognitive overload \cite{mccutchen2009writing} and that even small demands on working memory can lead to decreased fluency \cite{ransdell_structure_2002}. With this in mind, we hypothesized that an increase in writing variations could worsen this. As such, we aimed to design an editing interface that provides affordances for creating and comparing multiple writing variations without overwhelming the user.%

\textbf{Requirement 2: Support visually-structured management of variations.}
Documents can quickly become cluttered when trying to work with multiple variations of different text segments using current editing interfaces. Furthermore, the potential for LLMs to enhance writers’ ability to generate and revise multiple parallel variations of text further exacerbates the issue of clutter. Together, these factors highlight the need for a visually structured approach to managing  variations. Our goal was to support writers in seamlessly incorporating LLM-generated variations into the draft without cluttering the document or erasing existing content, and to retain the ability to simultaneously consider multiple variations. This is in alignment with prior studies that show that while users prefer being able to choose from multiple options \cite{dang2023choice}, having all of these options can negatively impact writing efficiency \cite{buschek2021impact}, indicating a need for option management support. Furthermore, having the ability to choose between options in an intuitive way can give users an opportunity to make more informed decisions as they compare and contrast between parallel ideas \cite{10.1145/1357054.1357074, 10.1145/1124772.1124960, buxton2010sketching}. 

\textbf{Requirement 3: Support context-sensitive variation comparison and revision.}
In a typical, linear document editing interface, we found it difficult to maintain a sense of the surrounding text as we created and added variations. This was particularly noticeable when trying to create and compare variations for smaller and embedded text segments--e.g., a sentence mid-paragraph or paragraph mid-section--which disrupted the text flow. Maintaining text flow is crucial since writers need to engage with information processing tasks such as ensuring the document maintains cohesion, which requires matching segments to the surrounding text \cite{mcculley_writing_1985}. This becomes increasingly challenging as the document holds more and more variations of different text segments. Our objective was to design an interface that allows writers to systematically generate and edit multiple variations in context of the preceding and subsequent text in the document.

\textbf{Requirement 4: Supports revision-centric, reusable, and non-linear LLM usage.}
Recognizing that revision is inherently nonlinear--with writers often revisiting earlier sections of a passage--and recursive, manifesting in repeated cycles throughout the writing process \cite{sommers1980revision, fitzgerald1987research}, we aimed to align our LLM integrations with this fluid, iterative nature of revision. Our goal was for writers to be able to use LLMs to manipulate text segments of varying lengths and refine them as needed in a way that is natural to their non-linear and recursive process. To enable this, we drew inspiration from design principles for visual interfaces, focusing on reuse, polymorphism, and reification \cite{10.1145/345513.345267}, and regarded LLM prompts as reusable, polymorphic commands that can be applied to targeted text segments of varying lengths. Recent research on designing LLM-powered writing interfaces has highlighted the value of viewing components of the LLM generation pipeline as interactive objects in supporting iteration and experimentation \cite{kim2023cells}. We adopt aspects of this approach in our design, such as reifying \cite{10.1145/345513.345267} LLM prompts into reusable AI Modifiers, and turning text segments into interactive Variation Fields akin to \textit{cells} \cite{kim2023cells}.

\subsection{Interface Elements}
We addressed these four design requirements by developing five interface elements using an iterative design process. To get the right design \cite{10.1145/1124772.1124960} the lead author iterated through multiple versions of each interface element, and tested them with five pilot users during in-depth brainstorming and design-review sessions over six months. 

\subsubsection{Design Process}

The design process involved transitioning from low to high-fidelity prototyping via three overlapping processes: (i) \textit{Paper-Prototyping} \cite{sefelin2003paper, snyder2003paper}: We began by considering an in-place editing interface as a foundational concept, and sketched various design elements that could be integrated into this interface to enhance exploration of variations; (ii) \textit{Cognitive Walkthroughs} \cite{lewis1997cognitive, mahatody2010state}: We used informal cognitive walkthroughs to rapidly iterate on our sketches and identify which potential solutions were intuitive; (iii) \textit{Web-based Prototyping} \cite{walker2002high}: We then developed functional versions of our intuitive design solutions in a high-fidelity prototype. There was considerable iteration between these stages. For example, if we saw issues in the high-fidelity, we would return to paper prototyping to devise refinements. Through this process, we developed five interface elements:

\subsubsection{Variation Fields}
\begin{figure*}
    \centering
    \includegraphics[width=\textwidth]{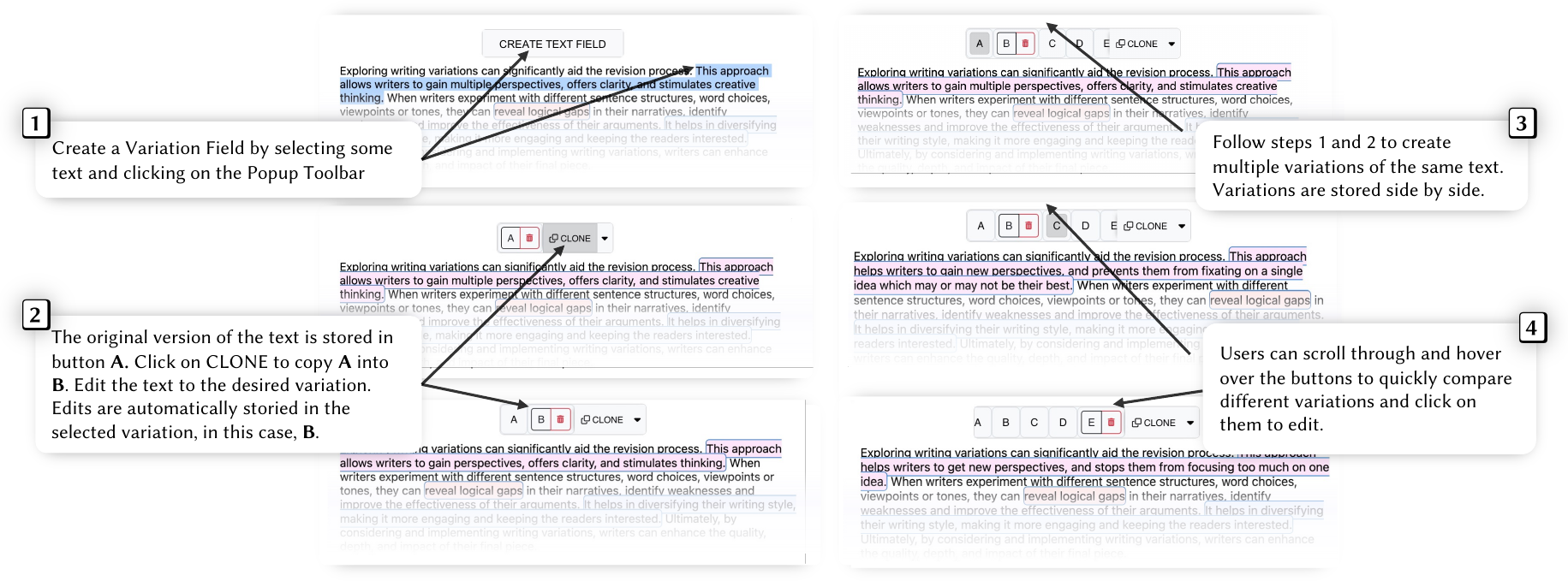}
    \caption{\textbf{Variation Fields \& Popup Toolbar:} Users can choose, clone and edit multiple variations stored inside flexible text fields.}
    \label{fig:hover_buttons}
\end{figure*}
Users can select any part of the text and create interactive Variation Fields that can hold multiple writing variations (Figure \ref{fig:hover_buttons}, Part 1). Newer variations can be added to an Variation Field without overwriting existing variations (Figure \ref{fig:hover_buttons} Parts 2 and 3) and without breaking the flow of the passage.

\subsubsection{Popup Toolbar}Variations are represented using buttons residing within a Popup Toolbar placed above the active Variation Field. The Popup Toolbar dynamically moves with the text caret, and users can peek at different variations by hovering over the corresponding buttons to view them in the context of the surrounding passage (Figure \ref{fig:hover_buttons}, Parts 4 and 5). Moving the cursor away from a toolbar button reverts the Variation Field back to the selected variation, allowing users to quickly compare between the selected and the hovered variations. Users can click on a button to select the variation and edit them in place, or discard it by clicking on the trash icon (Figure \ref{fig:hover_buttons}, Part 6).

\subsubsection{Variation Sidebar}
\begin{figure*}[ht]
    \centering
    \includegraphics[width=0.9\textwidth]{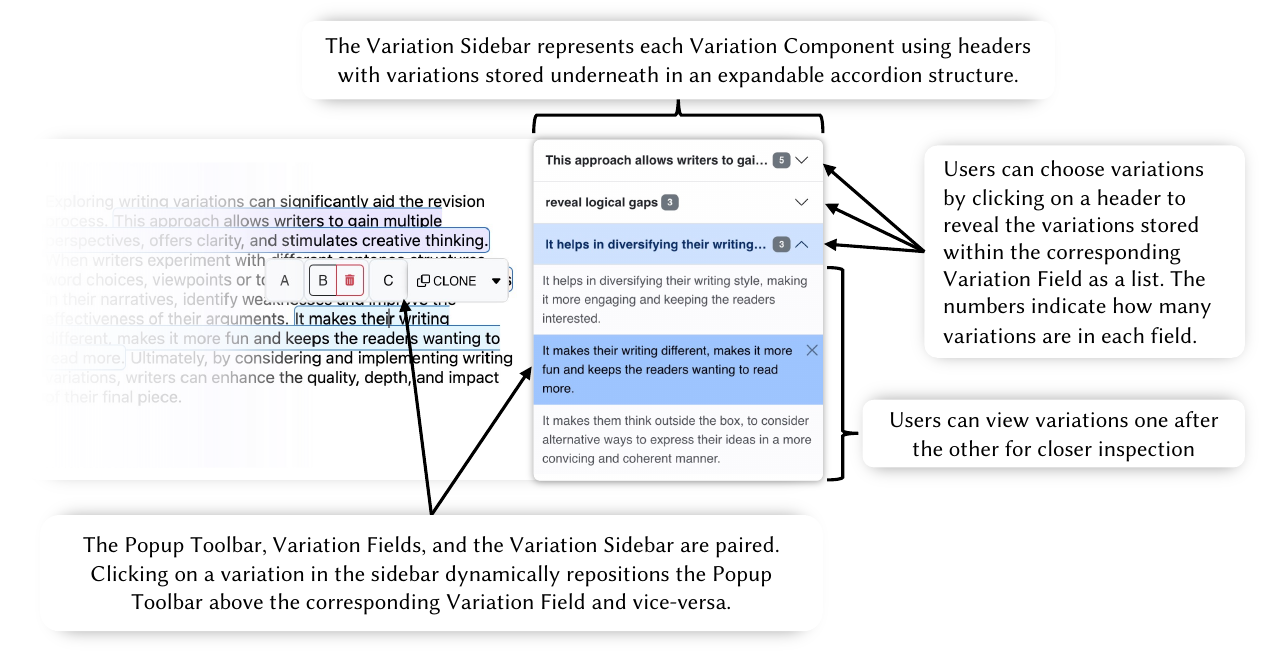}
    \caption{\textbf{Variation Sidebar:} Users can select, view, and navigate through multiple variations using an alternative accordion structure.}
    \label{fig:variation_accordion}
\end{figure*}
To help writers view multiple text variations together, and navigate through them more easily, we pair the Popup Toolbar with a sidebar organized in an accordion structure where each Variation Field has its own header, and all corresponding variations are stored underneath. Clicking on the variations in the sidebar dynamically re-positions the Popup Toolbar above the active Variation Field, and vice versa, allowing users to navigate through multiple variations in a visually structured manner (Figure \ref{fig:variation_accordion}). Users can show or hide the sidebar from the editor menu. 

The Variation Fields, Popup Toolbar, and the Variation Sidebar work together to address R1, R2 and R3. Note that the Popup Toolbar and Variation Fields have no AI capability, but support text revision by the user. To tackle R4, we developed two interface elements.

\subsubsection{AI Modifiers}\begin{figure*}[ht]
    \centering
    \includegraphics[width=\textwidth]{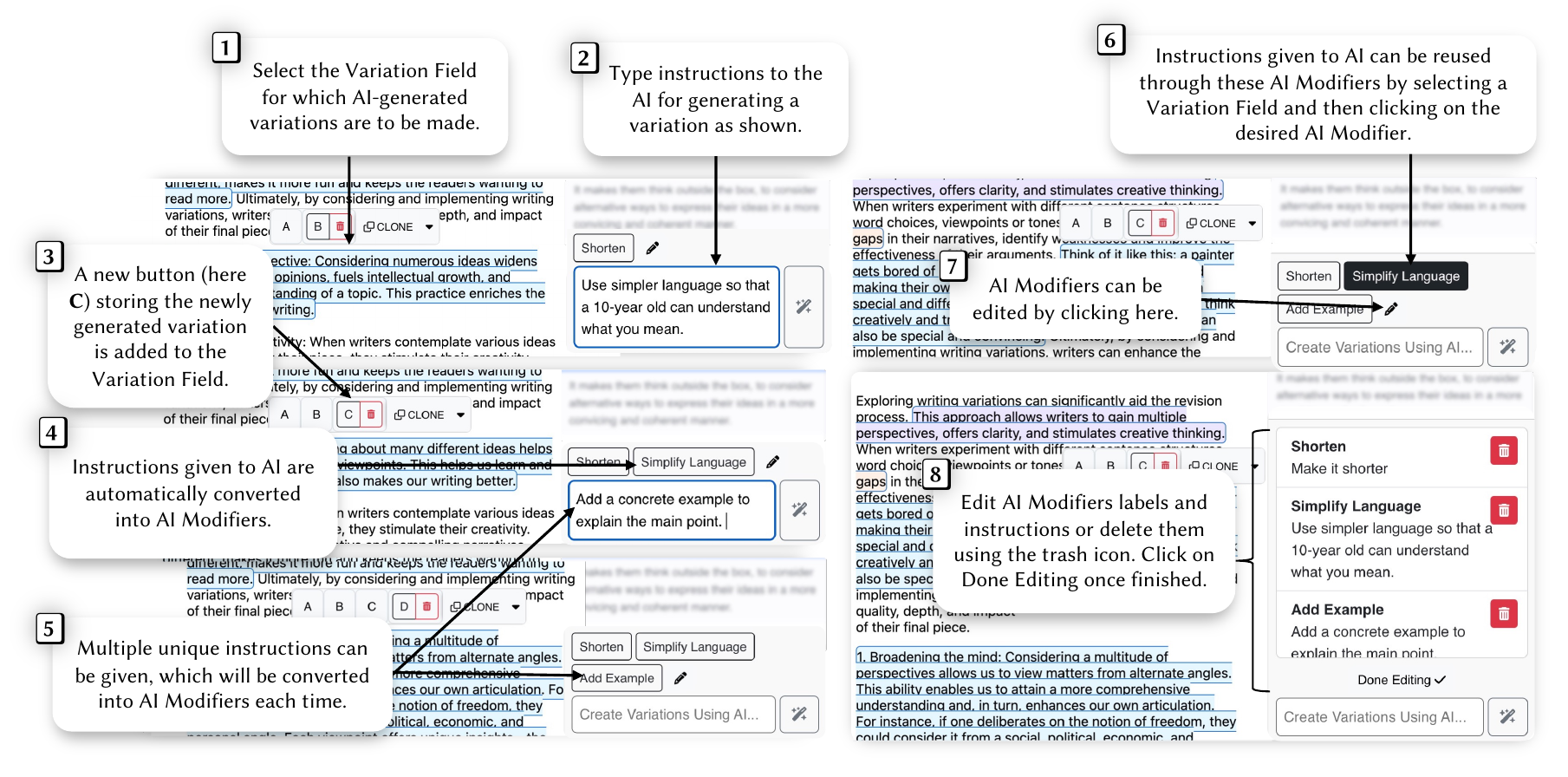}
    \caption{\textbf{AI Modifiers:} Variation Fields can also be edited using the AI Modifiers, which let users specify custom alterations. Descriptive labels are automatically generated for each AI Modifier to turn them into reusable buttons.}
    \label{fig:ai_buttons}
\end{figure*}
Users can generate new variations by selecting a Variation Field and typing instructions to the AI (Figure \ref{fig:ai_buttons}, Part 1 and 2). Instructions are automatically converted into labeled buttons (Figure \ref{fig:ai_buttons}, Part 4). The labels are generated using the LLM. As users experiment with newer variations using the AI, they create a set of custom AI Modifiers that they can apply to different parts of the passage, making them reusable (Figure \ref{fig:ai_buttons}, Part 5 and 6). The prompts and labels for the buttons can be edited and  improved over time (Figure \ref{fig:ai_buttons}, Part 7 and 8). This allows writers to not only create variations for a particular Variation Field, but also design buttons reflecting the kinds of variations they might want to generate for other parts of the text in the future, akin to a custom Swiss army knife for variations. As users add more variations, the interface organizes them into a scrollable list with a maximum height, so users do not run out of room. A more complete implementation could explore more sophisticated ways to group and search through AI Modifiers.

\subsubsection{AI Drafter}
\begin{figure*}[ht]
    \centering
    \includegraphics[width=\textwidth]{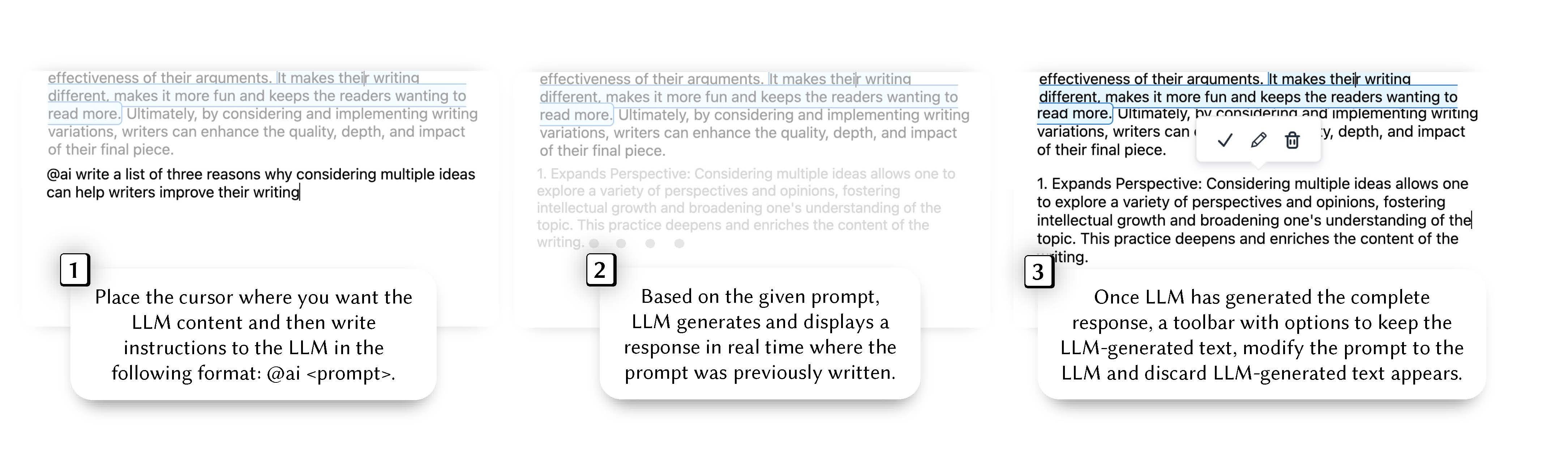}
    \caption{\textbf{AI Drafter:} Users can draft LLM-generated text directly into the document, providing tighter integration between the Human and AI generated writing workflow. Users can see the AI generated content in real-time and choose to insert or delete the output, or revise the prompt, giving them control over what is included in their document.}
    \label{fig:ai_insert}
\end{figure*}
Users can insert text from the AI model anywhere within the passage by prompting the system to invoke LLM usage in the following format: @ai <prompt>, and pressing enter. The text is generated and shown to the user in real-time, and they have the option to accept or discard the AI generated output, as well as revise the prompt to regenerate the output if it does not match what the user is looking for (Figure \ref{fig:ai_insert}).
\section{Evaluating ABScribe}
\label{evaluation}

To validate our design, we conducted a within-subjects evaluation study where we compared ABScribe to a carefully constructed baseline interface. Refer to \hyperref[baseline_appendix]{Appendix A} to see a screenshot. The Baseline interface featured the same rich-text editing capabilities as ABScribe without the ability to create and modify Variation Fields using the Popup Toolbar, Variation Sidebar and AI Modifiers. Instead, it featured a conversational AI assistant similar to ChatGPT, and the ability to insert AI generated text directly into the document as some modern AI editors such as Notion AI\footnote{\href{https://notion.ai/}{notion.ai}} have that capability. To minimize potential confounding factors arising from tangential differences between the study conditions, we maintained the same overall layout for common UI elements. This included the width of the sidebar, placement and dimensions of buttons and text fields, font size, and color. Additionally, we used the same underlying LLM model, GPT-4, to implement the generative AI features for both conditions. \\\vspace{-0.5em}

\begin{adjustwidth}{4em}{}
\begin{itemize}
    \item[\textbf{RQ1}:] How does ABScribe influence \textbf{user perceptions of the revision process} for AI-assisted exploration of writing variations when compared to the AI-integrated Baseline interface?
    \item[\textbf{RQ2}:] How does ABScribe influence \textbf{subjective task workload} for AI-assisted exploration of writing variations during text revision when compared to the AI-integrated Baseline interface?
\end{itemize}
\end{adjustwidth}

\subsection{Participants}

We recruited 12 writers (5 women, 7 men, aged 18 to 34) through flyers and study invitation messages on social networking services, such as LinkedIn and Facebook. All participants reported proficiency in reading and writing in English and were screened for prior experience in a broad variety of both fiction and non-fiction writing genres. We included participants with a range of prior experience levels in AI tool usage. This ensured that our findings were not limited to specific genres or AI usage patterns. All participants had ample writing experience, enabling them to effectively comment on the revision process and its relevance across various writing forms. Refer to Table \ref{table:participants} for writer profiles. Participants were compensated with 30 CAD. The institution’s research ethics board has approved the protocol.

\renewcommand{\arraystretch}{1.5}
\begin{table*}[ht]
\footnotesize
\begin{tabular}{ p{0.5cm} p{7cm} p{7cm}  }
\toprule
 ID & \textbf{Writing Experience} & \textbf{AI-Assisted Tools Usage}\\
\midrule
W1 & \emph{\textbf{Moderate:}} Worked as a staff writer for two political science publications. Writes fiction as a hobby. & \emph{\textbf{Moderate:}} Uses ChatGPT to edit writing projects, as well as receive feedback and suggestions for further passages, primarily after completing a passage to identify areas for further revision.\\

W2 & \emph{\textbf{Moderate:}} Taught ESL courses to non-native English speakers, specializing in IELTS, TOEFL, and business English instruction. & \emph{\textbf{Experienced:}} Uses ChatGPT by feeding it the main points of the article to generate a draft, and then editing the responses provided by ChatGPT.\\

W3 & \emph{\textbf{Highly Experienced:}} Has experience writing papers, specifically about writing tools for HCI. Has also, published a novel, and has publications in many highly regarded literary venues (BOMB, LitHub, FENCE, and more). Participant also writes their own music. & \emph{\textbf{Moderate:}} Uses ChatGPT for drafting messages, seeking feedback on fiction, and drafting small sections of research papers. Has experience with Respondable, a service in the Gmail plugin called Boomerang, for writing emails using AI.\\

W4 & \emph{\textbf{Highly Experienced:}} Writes fiction and published one novel, some sci-fi and fantasy short stories, and several articles for blogs, magazines, and satirical news sites. Worked as a staff writer as an undergraduate, a professional screenwriter for two independent studios. Also teaches two first-year writing classes in a liberal arts college. Achieved a MFA in Creative Writing. & \emph{\textbf{Limited:}} Briefly experimented with ChatGPT to test its capabilities by asking it to write some scripts, essays, and articles. Found the results to be amusing, but lacking in perspective and personality.\\

W5 & \emph{\textbf{Experienced:}} Writes content for social media profiles for an NGO. Studied English Literature during both bachelor's and master's degrees. Writes music, having penned 65 songs, and promotes it through social media and music platforms such as YouTube and Spotify. & \emph{\textbf{Experienced:}} Used Grammarly for on-the-go editing to write and ChatGPT much more extensively for both idea generation, as well as summation and synthesis of large bodies of text. Also found ChatGPT useful for helping figure out parts of creative works that may feel like they have gaps which can be prone to miss.\\

W6 & \emph{\textbf{Experienced:}} Worked for two national English language newspapers, including contributions to their weekend magazines, kid's sections, international section, in addition to also publishing fictional short stories for the newspaper. Completed a Creative Writing Certificate, and currently primarily writes about research. & \emph{\textbf{Limited:}} Used ChaptGPT in a very limited capacity, mostly to brainstorm assignment structures and topic sentences when writing.\\
 
W7 & \emph{\textbf{Experienced:}} Writes academic articles, general interest articles and reviews for local newspapers. Also writes short stories for sharing with friends, and has a short story published in an locally published anthology. Expertise is primarily in creative nonfiction. & \textbf{\textit{Limited}}: Briefly experimented with ChatGPT.\\

W8 & \emph{\textbf{Moderate:}} Written mostly technical papers, but also wrote some short stories as a hobby. & \emph{\textbf{Moderate:}} Uses ChatGPT every other day mostly to proofread, create templates for texts, and find the right creative direction when writing.\\

W9 & \emph{\textbf{Experienced:}} Usually writes research papers in computing education and blogs. Blogs usually cover personal experiences at work as well as hobbies. & \emph{\textbf{Limited:}} Used Grammarly and ChatGPT to edit writing.\\

W10 & \emph{\textbf{Moderate:}} Wrote some column articles for personal social media accounts, and several research papers over the past five years. & \emph{\textbf{Experienced:}} Uses Notion.AI, ChatGPT, and GPT-4. Uses Notion.AI for generating bullet points and brainstorming ideas, ChatGPT for generating templates for writing, and sometimes summarizing related work for research purposes.\\

W11 & \emph{\textbf{Highly Experienced:}} Focuses on academic writing such as papers, scholarship applications, reviewing, etc. & \emph{\textbf{Advanced:}} Has experience with Grammarly, Notion, Obsidian with GPT plugins, and ChatGPT. Mostly uses these tools to clean up sentences, and sometimes uses them to brainstorm titles for papers.\\

W12 & \emph{\textbf{Experienced:}} Engages in hobby novel writing, academic writing, blog writing & \emph{\textbf{None:}} Has not used AI-assisted writing tools, but has experience with AI image generation using written prompts.\\
\bottomrule
\end{tabular}
\caption{Self-reported experience with writing and using AI-assisted tools. Expertise labels for writing range from very limited to highly experienced, and labels for the AI tools ranges from none to advanced. Details on prior writing experience and AI tool usage is also included for each participant.}
\label{table:participants}
\end{table*}

\subsection{Tasks}

Each participant engaged in two guided writing tasks which were randomly paired with the two counter-balanced study conditions. Our choice of tasks—writing an email and a social media post—aimed to provide an ecologically valid writing experience that fits the timing constraints of our study and  served as realistic use-cases for LLMs, in alignment with recent HCI studies on human-AI co-writing \cite{kim2023cells, goodman2022lampost, buschek2021impact}, and commonly advertised applications of commercial AI-assisted writing tools, such as Grammarly Go, Respondable, and Copy.ai. We selected scenarios by considering situations that: (i) would be easy for users to imagine, such as emailing a professor or seeking a job as a copywriter; (ii) offer opportunities for exploring variations, such as devising multiple alternatives for a subject line or altering a sentence to maximize reader engagement. We wrote and tested the prompts to ensure the output was reasonable and of consistent length. For example, after initial rounds of testing, we found that the output generated by the AI was too long, and so we added `Keep it within three paragraphs' to both prompts. We asked participants to generate a draft using the same prompts for all scenarios. This approach was designed to keep their focus on exploring and organizing variations, rather than engaging in prompt engineering on the initial draft. It also helped maintain relative consistency in the generated output across participants and study conditions. Refer to \hyperref[scenario_appendix]{Appendix B} for additional details on the task scenario descriptions and prompts.

Then, with either ABScribe or the Baseline interface, participants were asked to use AI to explore eight variations (increasing or decreasing length, formality, word diversity, adding emojis, and two variations of their choosing) of three distinct text segments (title/subjectline, third sentence of the second paragraph, entire third paragraph), summing up to twenty-four variations. This approach ensured variation exploration of consistent number, size and variety across study conditions, while affording some scope for creativity as, for two out of the eight variations for each segment, participants had the autonomy to craft variations based on their preferences. We selected the variations by considering potential modifications writers might want to apply to their drafts, including changes in length, tone, and word choice. We then tested these modifications to ensure their feasibility with both ABScribe and the baseline interface.

\subsection{Measures}

To assess subjective task-workload, we used the widely used NASA-TLX \cite{hart2006nasa} procedure with weighting. To quantify the specific aspects of the LLM-assisted revision process that we aimed to improve, we also asked participants to rate their agreement on a 7-point Likert scale (1 = Strongly Disagree, 7 = Strongly Agree), similar to prior work \cite{lee2022evaluating, kim2023cells,lee2022coauthor}: 

\begin{enumerate}
    \item \textbf{Variation Granularity:} \textit{I felt like I could work with multiple (more than 5) writing variations of different fine grained parts of the text (e.g. word, sentence, paragraph) using this tool.}
    \item \textbf{Variation Search:} \textit{I felt like after creating all these variations, I could find previous variations when I needed them (e.g. when trying to create a new variation based on an existing variation I created earlier in the writing process).}
    \item \textbf{Prompt Reuse:} \textit{I felt like after creating all these variations, I could reuse my previous instructions or prompts to the LLM without having to rewrite them often.}
    \item \textbf{Variation Comparison:} \textit{I felt like I could identify fine-grained differences between multiple variations using this tool.}
    \item \textbf{Variation Editing:} \textit{I felt like I could systematically edit new variations without losing existing variations or cluttering the document using this tool.}
    \item \textbf{Variation Control:} \textit{I felt like I had control over which variations I wanted to keep, discard or change.}
    \item \textbf{Variation Divergence:} \textit{I felt like exploring multiple variations using this workflow will help me come up with variations that are surprisingly different.}
    \item \textbf{Draft Quality:} \textit{I felt like exploring multiple variations using this workflow will help me have better final draft.}
    \item \textbf{Intent Match:}\textit{I felt like exploring multiple variations using this workflow will help me come up with variations that are closer to what I want to say.}
    \item \textbf{Variation Diversity:} \textit{I felt like I could create variations with a lot of variability in word choice, style, and tone of voice using this tool.}
    \item \textbf{Document Clutter:} \textit{I felt like after creating all these variations, the document became cluttered.}
\end{enumerate}

\subsection{Procedure}

Participants began by signing a consent form and completing a survey that captured demographic data, prior writing experience, and familiarity with AI-assisted writing tools. The entire study lasted approximately 1.5 hours, and was conducted via video-conferencing software. Participants accessed the prototypes through their web browsers, mirroring how they typically access popular AI editing tools like ChatGPT and Grammarly. Conducting the study online enabled us to engage a diverse group of writers beyond Canada. After a brief introduction outlining the study's objective—to investigate the exploration and organization of multiple writing variations using LLM-integrated editing tools—participants undertook the two 15-minute writing tasks. Before starting each task, we demonstrated how the tools functioned and gave participants an opportunity to try them, ensuring they felt comfortable. After each task, participants completed the NASA-TLX assessment and 11 Likert-scale measures. These measures offered insights into the writers' perceptions of the revision process and prompted them to reflect on specific aspects of the process that we sought to improve through our design. The evaluation concluded with a recorded 30-minute semi-structured user interview on their experience with each interface. 

\subsection{Analysis}

Our data comprised interview transcripts, task observation notes, and the NASA-TLX and Likert-scale ratings for each condition. We coded and analyzed the interview transcripts and task observation notes using reflexive thematic analysis \cite{braun2019reflecting} through an inductive-deductive lens. The theory on revision-focused exploration of writing variations served as a pre-existing code guiding our interpretations. In a within-subjects design, we use pairwise one-sided t-tests to compare the sum of scores of NASA-TLX and our Likert-scale measures on the revision process. T-test was shown to be robust for aggregated data of this kind \cite{de2010five}. We aimed to determine if ABScribe presented significant improvements over the baseline interface, prompting us to select a one-tailed test with hypothesis $B < A$ for task workload and $B > A$ for level of agreement on the efficacy of the revision process. We performed an \textit{a priori} power analysis for a pairwise one-sided t-test, showing that we can detect an effect size of at least $d=0.8$ with $80\%$ power and a sample size of $N=12$ participants for a significance level of $\alpha=0.05$.
\section{Results}
\label{results}
\begin{figure}
    \centering
    \begin{minipage}{0.45\textwidth}
        \centering
        \includegraphics[width=0.9\textwidth]{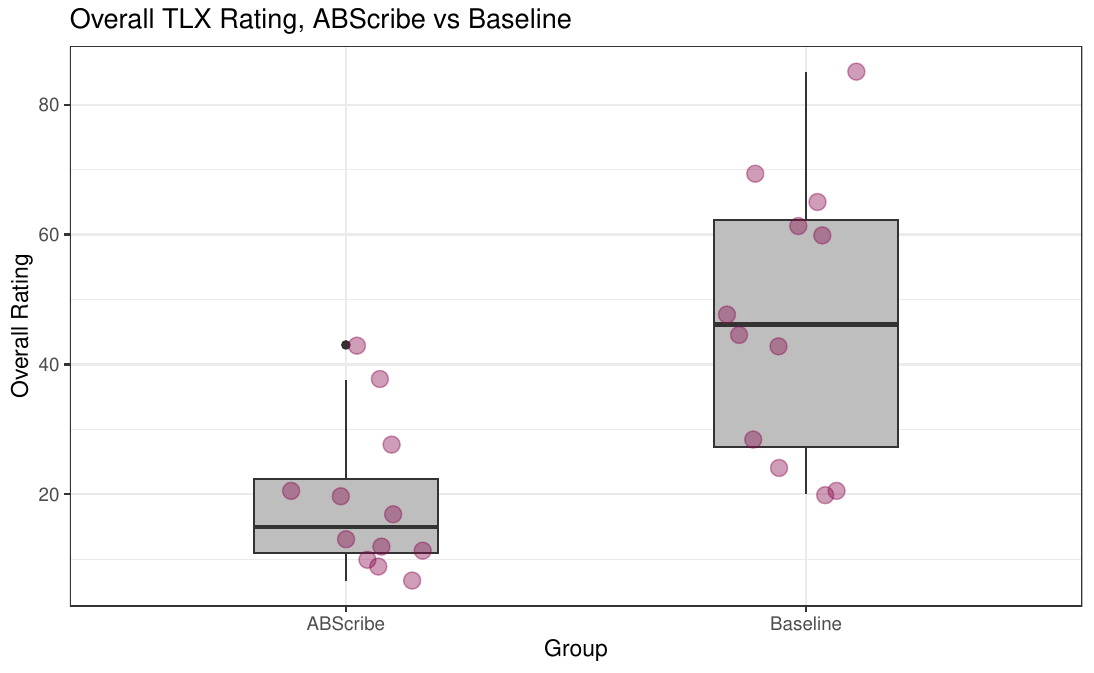}
        \caption{Overall results on NASA TLX subjective task workload}
        \label{tlx-boxplot}
    \end{minipage}\hfill
    \begin{minipage}{0.45\textwidth}
        \centering
        \includegraphics[width=0.9\textwidth]{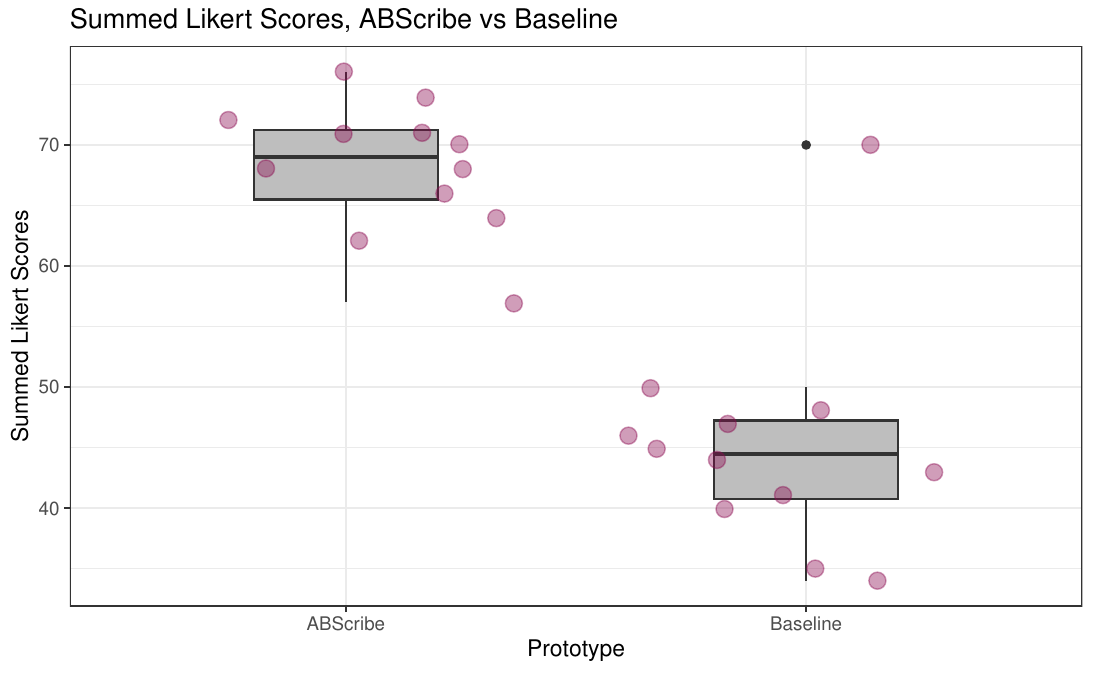}
        \caption{Summed Likert ratings for users' perceptions of the revision process}
        \label{likert-boxplot}
    \end{minipage}
\end{figure}

\begin{quote}
\textit{``The user interface for A [ABScribe] made comparisons, easy storage and access of those variations much easier than B [Baseline]. The fact that after you wrote a prompt, it instantly assigned a button to it that you could access later, was incredibly useful. It made it such that you could actually play with a more precise number of variations than I previously could and the fact that you could manually edit them and then again, quickly, have a way to play with the variation made it much more practical as a writing tool, there was a lot less physical effort involved to streamline that process. It was a wonderful, very smart way of dealing with the problem of clutter on the page. Even absent the AI generation part of it's an excellent way of storing variations.}'' -- W1
\end{quote}

\begin{figure*}
  \centering
  \includegraphics[width=\linewidth]{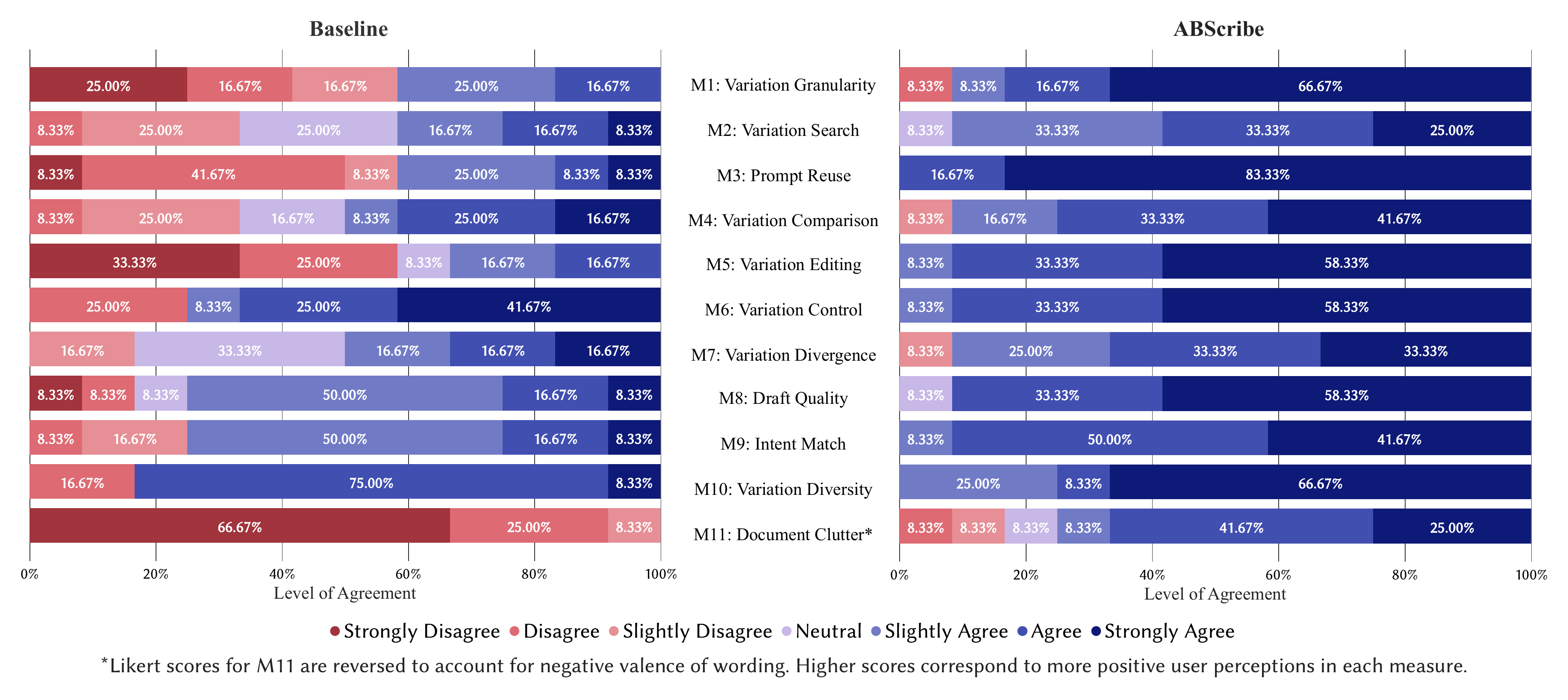}
  \caption{Responses to Likert-Scale Measure on the Revision Process for Exploring Multiple Writing Variations. Higher the agreement level, the more positive the user perception.}
  \Description{Figure illustrating the stress intervention design layout.}
  \label{fig bar}
\end{figure*}

The overall response to ABScribe, as exemplfied by by W1's comment and Figures \ref{tlx-boxplot}, \ref{likert-boxplot} and \ref{fig bar}, was positive, with a significant increase ($d = 2.41, p < 0.001$) in the summed agreement levels on the efficacy of the revision process (RQ1), and a significant reduction ($d = 1.20, p < 0.001$) in TLX rating for subjective task workload (RQ2) when compared to the Baseline interface. To gain deeper insights into the factors behind the reduction in task workload and the increase in user perceptions, we conducted a reflexive thematic analysis on the semi-structured user interviews. \textit{User perceptions on the revision process} (section \ref{rq-1-results}) groups findings relating to how the users perceived changes to the process and outcome of creating and managing variations as well as their interactions with the LLM. \textit{Subjective task workload} (section \ref{rq-2-results}) groups findings relating to the ease of specific tasks during the writing and revision process as well the ease of specific interactions with the interface to accomplish those tasks. 

\subsection{RQ1: User Perceptions on the AI-Assisted Revision Process}
\label{rq-1-results}

ABScribe provides features targeted at improving \textit{Variation Management} and thoughtful use of AI in text revision. To disentangle results of the two as well as identify when findings are a result of both those features, findings in the following section are tagged. We use \textit{Variation Management}, \textit{AI Integration}, or \textit{Mixed} to indicate whether a finding relates purely to the management of text variations, or how participants interact with the AI, or a mixture of both features. When a finding is \textit{Mixed}, comments are provided to exemplify what users attributed to \textit{Variation Management} elements, and what was due to \textit{AI Integration} elements.

\phantomsection
\label{f1}

\textbf{F1: Variation Management}--ABScribe lessens pressure to commit early to an initial idea and nudges users to explore a greater quantity of variations than the Baseline workflow. The non-linear approach to \textit{variation storage} within a \textit{Variation Field} and the ability to interact with them via the \textit{Popup Toolbar} and \textit{Variation Sidebar} without cluttering the document made some participants feel less pressured to commit to an initial variation before considering multiple options. In contrast, participants indicated that in traditional text editing settings, they maintained and iterated on a single variation. During the study, we noticed baseline interactions favoured expansion of variations for a single passage, revision, then collapse to a single variation to avoid clutter, whereas with ABScribe, multiple variations were maintained for considerations. For example, W3 notes how the difficulty of \textit{variation storage} in Baseline pushed them to maintain just one variation: \textit{``I would probably feel more pressured to just kind of work on one sentence and come up with a couple of variations and change it immediately. I would feel like I have to commit pretty early on, rather than generating a number of variations, trying a bunch of different tracks and sort of different timelines, seeing how each of them turns out and performing a master comparison at the end. So I think it had a significant effect, or would have a significant effect on my behavior, certainly doing the same task for both conditions where I was trying to deal with a bunch of different versions and trying to change them and revise them differently. It was vastly more difficult in B and a nonlinear approach seems to make much more sense.''}

W2, 3, and 12 specifically commented that the lack of interface clutter due to the way the \textit{Variation Fields}, \textit{Popup Toolbar}, and \textit{Variation Sidebar} handle \textit{variation storage} meant that they could create, revise and evaluate a larger number of variations with ABScribe. W12 noted that \textit{variation storage} in a linear flow is \textit{``just very clunky''}, leading to a \textit{``higher cost of  keeping multiple versions''}, \textit{``more scrolling in the document''}, and \textit{``taking a longer time to find anything specific''}. While exploring variations with Baseline, W12 noted that: \textit{``I sort of gave up about halfway because it was taking too long to vaguely remember, \textit{oh} there was this version that I thought was good, but I wasn't able to find it because the document became so long, so I just grabbed whatever to just finish.''}

\phantomsection
\label{f2}

\noindent
\textbf{F2: Mixed}--ABScribe enhances writers' ability to explore more granular variations in context of the surrounding passage. W1, 3, and 9 mentioned that they could work with smaller text-segments more easily when using ABScribe than Baseline. Participants indicated that this was due to both the Popup Toolbar enabling in-context comparison of variations (meaning they could switch between variations in-place observing how they would fit in to the passage) and the AI Modifiers allowing for less effortful use of AI on finer-grained text segments.

With regards to \textit{variation management}, a concern participants presented when editing fine-grained text segments within traditional text editing interfaces is that the overall coherence of the passage would be more difficult to maintain while performing \textit{variation comparison}. W9 said \textit{``I fear that if I edit small chunks, then the tone of different chunks end up becoming different. Whereas I kind of want to change the tone of the whole thing to one specific thing.''} W3 shared a similar concern but noted that the ability to view and edit variations within the context of the surrounding passage reduced their concern: \textit{``ABScribe is vastly superior for any kind of fine-grained edits which become incredibly difficult to deal with [in Baseline] if you're trying to do different edits on smaller variations of the text, unless you want to perform a single edit and immediate make that change.''}

With regards to AI Integration, When using the chat-based LLM interface in Baseline, W3 said they had \textit{``never even considered [editing] on a sentence level because it would be so hard to go into ChatGPT and say that in the third sentence of the third paragraph...I don't even know if it [ChatGPT] has a sense of where that is.}'' This sentiment was echoed by W9 who noted: `\textit{`Usually, [I] would be [editing] at least a whole paragraph and see edits from that and then paste some of those edits in [to the document] but I wouldn't put one sentence in...so there isn't very fine-grained control [in Baseline]...whereas here, [with ABScribe], because it takes less effort, I'm like okay, I can do one sentence.''} 

The combination of AI Modifiers and Popup Toolbar enhances writers' ability to explore more granular variations by enabling the writer to remain within-context as they edit. 

\phantomsection
\label{f3}

\noindent
\textbf{F3: AI Integration}--ABScribe nudges writers toward an imperative LLM prompt writing style in contrast to the conversational style in Baseline. W1, 3, 6 and 8 noted how the automatically generated \textit{AI Modifers }that captured prompts they previously wrote influenced their prompt writing style to be more direct. W1 mentioned that in ABScribe, \textit{``you're conscious of the fact that you are designing a button, and that forces you think within that framework. You're not really asking someone to do something [like in the chat-based interface]. You're giving it instructions to make a button...which make things very simple, very easy, very quick...making it much easier to create prompts, and then use them subsequently, when they're being instantly converted to a new use-case.''}  Some examples of the short imperative style from the participants include \textit{``Make more professional, Make more casual, write this so a fifth grader could understand it''} Note the imperative style and general nature of the prompts.

In contrast, the conversational approach sometimes led users to anthropomorphize the AI, using superfluous words that do not influence the quality of the variation generated. For example, we observed that W8 would say ``\textit{Can you please make this variation shorter?''} or ``\textit{Could you make a version with more varied vocabulary''}'' when instructing the chatbot, but opt for a more direct \textit{``make it shorter''} prompt for ABScribe. When asked why, they said the conversational AI assistant felt \textit{``similar to Clippy''}, referencing the Office Assistant from a discontinued intelligent interface from Microsoft with an interactive animated character \cite{Baym_Shifman_Persaud_Wagman_2019}. 
\phantomsection
\label{f4}

\noindent
\textbf{F4: AI Integration--ABScribe nudges writers toward composing generalizable and atomic LLM prompts in contrast to the complex and variation-specific prompts in the conversational approach in Baseline.} W3 and 12 shared that with ABScribe, they were intentionally trying to simplify their prompt design to make them generalizable and reusable across different text segments. W3 said \textit{``I felt more like I was going to create a generalized prompt, that I will probably reuse later. So it felt like it ought to be something that's simple that could be applied to a variety of situations rather than something that's specific to a single piece of text.''} In contrast to W3, W9 found that due to the ease of prompt reuse, they were more likely to create longer prompts that would generalize to other contexts, when compared to the chat interface. \textit{``I was okay with writing longer prompts, for example, to imitate the style of a character because it took less effort, and it was fun for me to do, and I knew I could use it again in other sentences. Whereas in [Baseline], I wasn't as excited because it would take more time to type, and I knew that if I had to reuse it, I would need to type it again.''}

However, because participants were actively thinking of instructions they could reuse across different text segments, they were less likely to write prompts that were tied to specific characteristics of particular text segments, as exemplified by W3's comment: \textit{``I categorically preferred A [ABScribe] to B [Baseline]...the only advantage to B might be that it slightly encouraged me to have more nuance in the prompts I gave to ChatGPT...it just made me realize I could do that. I know I could do that in ABScribe too, but the button interface kind of guided me to more naturally consider simplistic prompts rather than prompts which might be more suited to particular tasks for those in conversation with a bot.''} 
A prompt pattern in the conversational approach was to copy paste a section of text then ask for a modification such as the following by W3 `\textit{`"Comprehending the essence of crisp, contemporary material ..." Can you write this in a way that maximized the number of linkedin likes?}''

With ABScribe, participants tended to decompose specific prompts that only apply to a given text segment into more general, atomic prompts that apply across variations. To achieve similar specificity, participants \textit{combined} multiple prompts by clicking selecting multiple AI Modifiers. W3 considered mimicking the functionality of longer, more complex prompts by \textit{``stacking''} multiple of their general, more atomic prompts to explore variations: \textit{``I think I would also be more willing to try out a variety of permutations of different prompts, rather than trying to apply one prompt globally to an entire email, or full prompts to one or two sentences. So seeing how compound prompts might help at various points would be a lot better in A [ABScribe].''} W8 demonstrates this stacking of atomic prompts in the following use of the AI Drafter ``\textit{@ai tell me a joke, make it more creative, personalize!, add a cough or a speech utterance between every few words}'' Note the use of several generic and atomic prompts "make it more creative" followed by "personalize!"

\subsection{RQ2: Subjective Task Workload}
\label{rq-2-results}
Participants pointed to four aspects of the revision workload in ABScribe that led to the significant reduction in subjective task workload: reduction in clutter, ease of \textit{variation management} which referenced both \textit{variation storage} and \textit{variation comparison}, access to surrounding text context, and the reduction of context switching during LLM use.
\phantomsection
\label{f5}

\noindent
\textbf{F5: Variation Management}--ABScribe reduces task workload by reducing document clutter. W4, 5, 7, 10, 11, commented that the general lack of clutter in the ABScribe interface when dealing with multiple versions affected various aspects of their writing and revision process. Specifically, the way ABScribe handles \textit{variation storage} in-place (i.e., all variations of a segment of text are stored and can be accessed at the \textit{Variation Field} using the Popup Toolbar) as opposed to in-sequence, by storing both versions in the main text editing interface, reduced clutter. W11 summarizes this in the following quote \textit{``I think definitely the biggest difference would have been the fact that your your document is not as cluttered. Actually, it doesn't get cluttered basically because you can just switch between versions and the text is on in the same location. So that's already a huge boost in terms of not having a mess on my hand. And you notice at the very start I was already organized to think that way.''}
When comparing ABScribe to the linear storage of variations (one after the other within the main document) in Baseline, W5 and 8 noted that the linear interface inevitably led to cluttered and messy documents. W5 pointed out the difficulties of managing variations in the linear Baseline approach: \textit{``So the linear approach was more difficult, because there was just walls of text, like they began piling up very quickly. And I tried to segment them, right, I believe, if I numbered them, it would have been better. But at the same time, it doesn't get rid of the root problem, where more and more text is, is being added to the whole draft.''}

A noteworthy and surprising trade-off is that although all participants agreed that the ABScribe interface produced less clutter, two of them perceived clutter as unproblematic or even beneficial. W2 and W6, were both comfortable with the workflow afforded by Baseline as it was similar to what they were used to doing in their naturalistic writing tasks - paper, article, and book writing. They noted that clutter didn't matter as much to them during the revision process. W2 mentioned that Baseline \textit{``definitely''} felt more familiar, and that \textit{``...you probably think cluttering is one of the one of the important factors to consider when people are writing, I actually don't think that's the case. That's why I don't care about whether it looks [messy during revision].''} W2 also mentioned when doing \textit{variation comparison}: ``\textit{In terms of comparison, I think this is just, I don't know, it just, it looks convenient. But actually, when you're actually writing things that matter you don't want to just use this to compare the sentences. ... For example, I guess most people use ChatGPT to write emails, that's the most popular application, when they write emails to clients and professors, people—they don't want to offend them. But sometimes you want to pay attention to your wording and stuff, in that case, I don't think anyone wants to just compare like this. You want to be careful.'' and concluded with ``I would not compare them on this page [via the Popup Toolbar and hover feature], I would compare them in the Idea Bucket [Variation Sidebar].}''

W6, who described themselves as a \textit{``a messy editor''}, a \textit{``hoarder''} of various made copies of older text, felt \textit{``really conflicted''} about the reduction in clutter. They commented: `\textit{`I feel like it would be a little bit easier to do that [variation comparison], because the text is already there [in the main document editor] ... you can't see two versions side by side unless you were to scroll in the box [Variation Sidebar], right. And so I feel like version B would be just like, it's easier there to do it. [variation comparison]}'' and that they liked being able to \textit{``mishmash multiple versions together''} in a messy document. In describing the workflow in Baseline, they said it was \textit{``more familiar to me than like doing it the way that you would in ABScribe. Even though in an abstract way, the non-linear approach makes a lot of sense...I feel like it just feels like that's how the design should be...but I feel like [I'm] a messy editor. And so it's, it's almost easier for me to edit, in Baseline.''}
\phantomsection
\label{f6}

\noindent
\textbf{F6: Variation Management}--ABScribe reduces task workload by enhancing variation management. Two major activities in the exploration of multiple variations were \textit{variation storage}, or tracking \textit{variation history}, as W7 referred to it, and \textit{variation comparison}. W3, 6, 7, and 11  noted that ABScribe's non-linear storage of variations necessitated less overhead to track \textit{variation history} and do \textit{variation comparisons} . 
W3 commented that by having the ABScribe interface manage \textit{variation storage} for them as opposed to manually placing text variations in the document, they were more able to focus on the writing task: \textit{``You're fully focused on the writing changes you're trying to make, as opposed to managing the state and managing your document and managing like, that kind of stuff. So that was the biggest difference for me. So I really liked that feature. And that made a huge difference in general to the task. But because I'm less focused on management of things, or management of my thoughts a little bit, it's a lot easier just using that, like the versioning system.''} 

While W3 touched on the ease of \textit{variation storage}, W11 discussed how the Popup Toolbar enhanced the ease with which they did \textit{variation comparison}: \textit{``the feature made it easier to to do the comparisons, because then you can click the version that you're - you're comparing with, and then hover and look at the text [for the other variations]. Whereas, with the Baseline, you have to both keep track of where the version you're comparing with is and also simultaneously figure out which version you're comparing to...so that is trickier than dealing with the new approach.''}

A trade-off some users commented on was that it was easier to compare large segments of text in-sequence within the main document as is done in the Baseline. W3 and W7 both noted the benefit for comparing variations in-sequence within the document. W3 said ``\textit{Yeah, I mean, one nice thing about Baseline is that I do get to see all of the versions together [viewed at the same time in the main text editing space]. So they're all listed for me}'' and W7 stated ``\textit{I think in that particular scenario[taking a small part of  variation 1 and merging it with most of variation 2 in order to make my best variation], having a linear approach [like Baseline] may be more helpful for me to like, see all the different variants and then compare?}'' 
However, they both followed their statements by noting the \textit{Variation Sidebar} could serve a similar purpose so the trade-off was perceived as advantageous for ABScribe. W3 commented: \textit{``[with Baseline] it's much harder to have a number of sentences, which you're generating different versions for because even once you start to hit two different sentences, [and] we're trying to generate different variations, the document becomes very cluttered, and becomes difficult to manage. And you forget what the context is for each of those different variations. So for context, and for clutter, I think A is vastly superior.''}  and W7 said ``\textit{your idea bucket [Variation Sidebar] does afford us that feature as well. Because I found myself doing that a lot. I was checking when, when we were first testing out the tool and giving me comps, I was checking to see if it was different enough. And we weren't I wasn't constantly like toggling in the document text, Right? So I was just like scanning the idea buckets [Variation Sidebar], I don't think that you're losing out on that feature, necessarily, just because I'm seeing that the linear approach is more helpful}''
\phantomsection
\label{f7}

\noindent
\textbf{F7: Variation Management--ABScribe reduces task workload by showing variations in context of the surrounding text during manual editing and when editing with AI.} The need for considering context during the revision process came up during several interviews. The benefits of \textit{variation management} on editing variations with context are exemplified by W3-8, and 10's comments. They note that thanks to the in-place\textit{variation comparison}  via the hover interaction with the Popup Toolbar, they were able to see what a variation looks like within a paragraph. To this effect, W10's states: ``\textit{I like the nonlinear version, because when you hover on the different buttons, you can directly see the impact of different variations within the paragraph or within the context. So in that way, you know, whether the text fit into the original document or not. Whereas in Baseline, if you put all the variations linearly in the document, at some point, you just start to lose a sense of what's the context of this of this sentence, what am I writing there?}"
W7 echoes the value of viewing variations with surrounding context: \textit{``When you're writing a paragraph, you're not looking at a sentence in isolation. So if you're changing a particular sentence, you want to see how it looks in comparison to the rest of your text. And so to have the nonlinear version allows you to kind of do that more seamlessly than with a linear version, where you'd have to reorganize a lot more in order to have that effect.'' Whereas, W10 mirrored this, and noted that manually organizing variations while simultaneously figuring out context was challenging: ``I need to manually think of a way to organize all the variations so that I understand what they mean. Or like, what, how they're connected to the original text. That cost a lot of time. And it's like, very high physical demand.''}

\phantomsection
\label{f8}

\noindent
\textbf{F8: Mixed--ABScribe's variation storage and in-place LLM revision reduces subjective task workload.} Interactions that bring the user outside of the primary text editing interface and break their flow of writing or revising were perceived to be effortful and time-consuming. 
W7 points out how \textit{variation management}--particularly \textit{variation storage} - with the Baseline interface necessitates context switching: \textit{``So if I were to go into version history, and I wanted to go back to a very particular change in one particular paragraph, but I made that like 50 changes ago, I would either have to revert back to something where all of the document would have been unchanged, or I'd have to do like a very inconvenient and kind of cumbersome process of like copying that particular change from that particular version history into my current doc and then proceeding, which is, like tedious''} 

W4, 5, 7, and 9  commented that they all had various ways of handling \textit{variation management} in their naturalistic writing tasks such as producing articles, research papers, and books, which were similar to how they handled variations with the Baseline interface. These methods, such as copy-pasting different versions from separate documents into the primary text editing interface for comparison in-sequence, or doing \textit{variation comparison} side by side in separate windows required them to leave the primary text editing interface to perform \textit{variation comparison}. W5, 7, and 9 all found \textit{variation comparison} less cumbersome with ABScribe, especially when performing edits on several smaller text segments, due to the lack of context switching. W9 explains this here: \textit{``Whereas here [ABScribe], I think because it takes less effort, like okay, I can do one sentence. I also want to do another one. So I'll do that. I don't need to copy the whole paragraph in [the chat-interface of Baseline] and try to get an answer from that. I can just do it to those sentences.''}

Use of AI within the Baseline was also indicated as a source of context switching. W10 notes how the chat-like interface in Baseline pulls them out of context: `\textit{`Also, the chatbot in Baseline, I'd say it's pretty much [the] same as ChatGPT. So if I want something, I need to scroll back to try to look for it. So that's pretty much similar to the current AI writing system.''}  ABScribe's AI modifiers and AI Drafter enable the use of AI without requiring leaving the main text editing interface.

\phantomsection
\label{f9}

\noindent
\textbf{F9: AI Integration-- ABScribe reduces task workload by making prompts more reusable than Baseline.} Almost all users (W2, 3, 6-10, 12) noted that prompt reuse was much easier in the ABScribe interface. W3 commented on how the Baseline interface imposed a \textit{``memory load or cognitive load issue to remember what prompts you have.''} W2 notes how the AI Modifiers of the ABScribe interface eliminate the need to rewrite prompts stating \textit{``I don't need to rewrite prompts every time. It was really very quick and efficient. The usability of this one in terms of buttons, the reusing the prompts is very good.''} While W8 called ABScribe's AI Modifiers \textit{``much more streamlined.''} W3 sums up the interaction concisely in the following quote: \textit{``A is vastly superior for reusability, there is no question. B, you basically have to work from memory, which can also be fine. But with version A, you click it, you don't have the same memory load or cognitive load issue to remember what prompts you have before, cannot be more better facilitated.''}

\section{Discussion}
Writing plays a crucial role in many people’s lives, and the interfaces used for editing text can deeply influence the writing process. As LLMs increasingly blur the boundaries between human and AI writing capabilities \cite{zhou2022large, grace2018will}, the HCI community needs to explore interfaces that improve  writers' interactions with AI. In our work, we contributed an ensemble of five interface elements that significantly enhance user perceptions on AI-assisted revision (RQ1) by supporting rapid exploration and organization of multiple writing variations without overwhelming users (RQ2). 
\subsection{Design Implications}

To interrogate the broader applicability of our results toward other contexts, and to characterize how ABScribe may relate to creativity support tools/authoring applications in general, we consider the following design implications: 

\noindent
\phantomsection
\textbf{D1: Emphasize Parallel Storage of Text Variations To Nudge Users Toward Working with a Larger Number of Increasingly Granular Ideas} In line with HCI design principles that highlight the importance of investigating multiple alternatives \cite{buxton2010sketching, 10.1145/1124772.1124960}, we have developed ABScribe to facilitate the rapid exploration and organization of many text variations, which is a problem exacerbated by AI. Our findings suggest that parallel storage is a useful approach for inspiring writers to generate numerous ideas for granular sections of text, without becoming overly attached to any single variation. Although our implementation deals with text-segments, the challenge of organizing proliferating ideas through parallel storage extends beyond textual content, to other forms of creative content. As companies advance foundational models for manipulating various types of inputs — such as Google's recently announced `natively multimodal' Gemini model \cite{Pichai_Hassabis_2023}, described as being adept at `reasoning seamlessly across text, images, video, audio, and code' \footnote{\href{deepmind.google/technologies/gemini}{https://deepmind.google/technologies/gemini/}} — designers can explore ways to extend our design to creativity support tools for other modalities. For example, an application for editing AI-generated images could implement an AI Modifier which accepts textual instructions and generates corresponding image variations, which are then organized within Variation Fields, contributing to HCI research on text-to-image AI editors \cite{brade2023promptify, 10.1145/3544549.3577043, liu2022design, oppenlaender2022creativity, wang2023reprompt}. The Popup Toolbar could be adapted to non-linear layer management in image editors \cite{chen2011nonlinear, barrett2002object}, or applied in presentation software, allowing users to experiment with different headings and bullet points on a slide.

\noindent
\phantomsection
\textbf{D2: De-Emphasize Parallel Storage to Support Editing Forms that Are Better Suited to Linear Representations of Content.} Our findings also suggest that while parallel storage nudges users towards exploring multiple text variations, it remains important to cater to linear editing preferences. Many participants preferred to view previous variations side-by-side or in-sequence when producing a new variation. For example, W6 noted their habit of linearly organizing text variations for side-by-side comparison, which seemed less intuitive in ABScribe. Although our design does not \textit{prevent} the sequential pasting of text blocks, it subtly \textit{discourages} linear editing. This insight underscores the potential benefits of a Variation Sidebar as a supplementary tool for linear viewing of variations, complementing the Popup Toolbar's parallel functionality. A more complex Variation Sidebar that allows for reordering and editing inside the accordion could help with linear revisions. The Variation Fields could also be changed to allow for both parallel and linear editing, such as through incorporating an 'explode/contract' button in the Popup Toolbar that allow users to switch between a linear, exploded display of variations for easier comparison and a more compact, collapsed view within the Variation Field. Together, D1 and D2 contribute to existing discussions surrounding linear versus non-linear representations of versions history across various domains \cite{head2019managing, o2009making, chen2011nonlinear, plaice1993new, zhang2023vrgit, tichy1982design}.

\noindent
\phantomsection
\textbf{D3: Scaffold Prompts Within Discrete Interface Elements to Encourage Action-Oriented Prompt-Writing.} If the goal for designers is to guide users towards crafting more succinct and direct prompts, incorporating scaffolding around specific interface elements signifying actions can be effective. This approach is evident in the buttons created by AI Modifiers. In contrast, in free-form conversational user interfaces, we saw a tendency for users to adopt a more polite and verbose style. For example, participant W8 would say, “Can you please make this variation shorter?” accompanied by a pasted text snippet, rather than simply selecting a Variation Field and requesting, “Make it short.” Of course, there are contexts where encouraging a polite and detailed interaction with AI agent is beneficial, e.g., in training scenarios where a customer service representative practices dialogues with an AI. However, in situations focused on productivity tasks, such as composing an email, steering users towards a direct prompt-writing styles can help. Our findings illustrate how interface elements can noticeably influence prompt-writing style, akin to previous work on, such as Dang et al.’s investigation of diegetic and non-diegetic prompts \cite{dang2023choice}. It also adds to prior work using objects to represent AI prompts \cite{kim2023cells} and text segments \cite{han2020textlets, 10.1145/345513.345267}. 

\noindent
\phantomsection
\textbf{D4: Reduce User Effort by Turning AI Prompts into Customizable Toolkit for Digital Tasks.} AI Modifiers in our interface highlights the potential of LLM prompts as tailored sets of reusable user actions, resembling a bespoke Swiss Army knife. For instance, users who frequently need to condense, translate, and proofread text can effortlessly create AI modifiers for these tasks, facilitating reuse and minimizing effort. Previous studies have shown that users may hesitate to write prompts when given multiple AI suggestions \cite{dang2023choice}, underscoring the importance of reducing prompt-writing effort. Our design addresses aspects of this problem.

Unlike traditional menu buttons which have predictable outcomes, AI Modifiers can yield unexpected results due to the inherent unpredictability of AI models \cite{yampolskiy2020unpredictability, yampolskiy2019unpredictability, holland2020black,hauhio2023spectrum}. Prior research on AI-assisted editing suggests that users may find imperfect AI outputs useful \cite{10.1145/3490099.3511105}, given that they have the ability to edit. Thus, while using AI prompts as personalized toolkits is beneficial, designers will have to ensure that users can easily refine both the AI’s imperfect output and the unpredictable instructions encapsulated by the AI modifier.

The current implementation of AI Modifiers as labeled buttons is only one of many possible approaches. Alternative user experiences could include AI-generated icons, options within a nested menu, or even simplified voice commands, for example, a user might initially issue a detailed voice command and then receive a suggestion for a condensed version for future use. Our design could be also be extended to give users more control over AI features, such as adjusting the 'temperature' to modulate predictability. 

\subsection{Limitations and Future Work}

A noteworthy limitation of our work is that we constructed the baseline interface rather than opting to use an existing system in our evaluation. We chose to construct the baseline because early pilots of our evaluation study revealed that utilizing a real-world system as a baseline (in our case, ChatGPT and GoogleDocs) introduced several logistical challenges that made it difficult to control and run the study. For instance, to ensure ChatGPT used the same underlying LLM model, we had to provide participants with login details to a paid ChatGPT account, potentially exposing credit card details. When using Google Docs, participants would resort to using editing features outside the focus of our research, such as user comments. Our choice of constructing a baseline is in alignment with prior HCI research on AI writers \cite{10.1145/3490099.3511105}. However, this may have affected the ecological validity of our results. Future studies could compare ABScribe with real-world AI-assisted writing workflows.

There are also limitations to external validity common in lab-based evaluation studies. First, our evaluation study was restricted to English writers. Although the underlying LLM, GPT-4, supports multiple languages \cite{gpt-4}, our study focused on English writing tasks. Therefore, we would caution designers against directly applying our design without further evaluation. Second, our evaluation study was limited to a single 1.5-hour session with two guided writing tasks to ensure we could conduct our comparison in a controlled setting. Future work could investigate how writers select their own writing task and revise text over extended periods, possibly spanning several days.

A potential risk with any generative AI tool is that the AI may produce \textit{hallucinated} or \textit{confabulated} content \cite{salvagno2023artificial, emsley2023chatgpt, smith2023hallucination, ji2023survey}. ABScribe's AI Drafter and AI Modifiers both present a risk of introducing hallucinations into the text, which becomes especially important to be cognizant of when writing in high-stakes domains or in educational settings. We carefully named our features to account for this. With the AI Drafter, the tool is specifically referred to as a \textit{drafter}, not a \textit{writing companion}, to nudge the user toward thorough revision of what it produces. The AI Modifier rephrases existing text written by the author, which provides the generative AI with a baseline grounded in the author's text, which is expected to be fact-checked.

Finally, it would be useful to implement logging features into ABScribe to concrete measures from the tool, such as: How many words did the subjects actually write; how many words were written by the LLM; what percentage of the resulting text was from the LLM vs. the human; how many times did the subjects interact with the various features in the ABScribe UI? We did not collect these metrics as our focus was on addressing task workload and user perceptions. Future studies could investigate the relative contributions of human and AI in co-written documents.

\section{Conclusion}

In this work, we presented ABScribe, a human-AI co-writing interface built for swiftly exploring multiple writing variations using Large Language Models (LLMs). Our design only markedly decreasees task workload $(d = 1.20, p < 0.001)$ but also bolsters user perceptions of the revision process $(d = 2.41, p < 0.001)$, in comparison to a baseline chat-based AI editing workflow. We evaluated our design with writers (N=12) to validate the efficacy of ABScribe and offer insight into how writers leverage LLMs to explore variations, revealing a preference for non-linear over linear revision strategies, especially when engaging with a multitude of variations at finer granularity levels. We also found that scaffolding LLM use with task-focused UI components, like buttons, encouraged writers to create more generalizable prompts and use more direct, imperative language in prompt design. Our work informs HCI research on the design of Human-AI writing interfaces for the rapid exploration of writing variations. 
\begin{acks}
We are grateful to the writers who participated in this work. We would also like to thank Siqi (Jessica) Zhang and Divya Gupta for their assistance with the transcription of interviews, and Steven Tohme for his help in open-sourcing the codebase. This work was partially supported by grants from the Office of Naval Research (N00014-21-1-2576), the Natural Sciences and Engineering Research Council of Canada (RGPIN-2019-06968), and the National Science Foundation (2209819).
\end{acks}

\bibliographystyle{ACM-Reference-Format}
\bibliography{references}


\appendix
\section{Baseline Interface}
\label{baseline_appendix}

\begin{figure*}[!b]
    \centering
    \includegraphics[width=\linewidth]{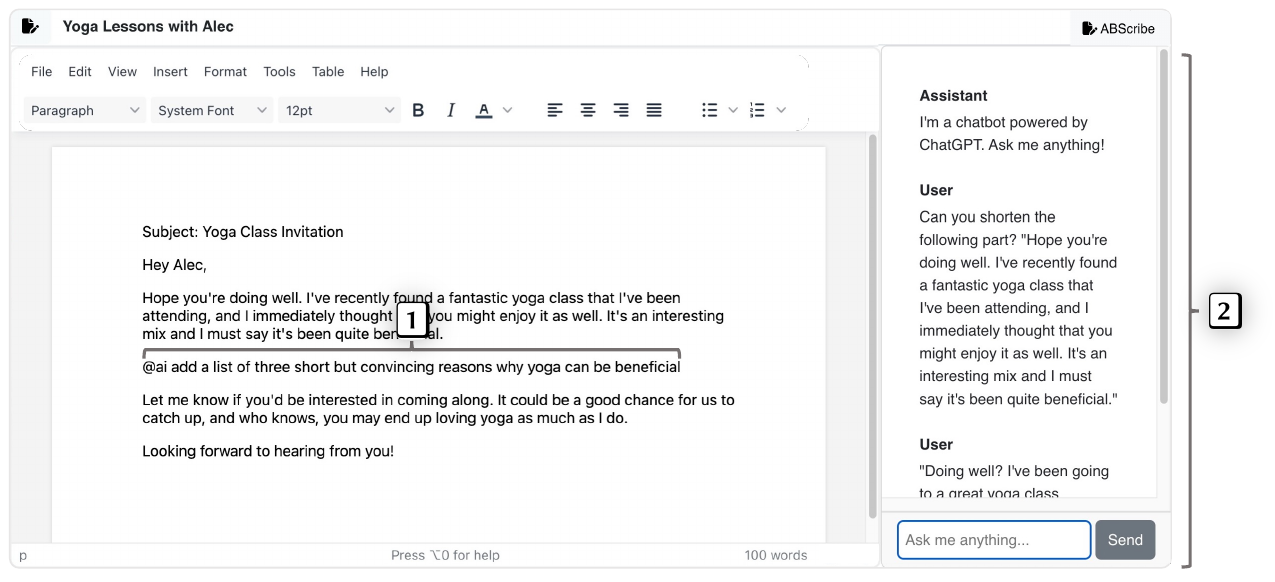}
    \caption{\textbf{The Baseline Interface:} (1) We retained the ability to insert text directly into the document using AI Drafter as some modern AI editors have similar capabilities. (2) The chat-based interface on the left used the same underlying model (GPT-4) as ABScribe. All tangential differences such as font size and rich-text editing capabilities were consistent with ABScribe.}
    \label{fig:baseline_interface}
\end{figure*}

The Baseline interface featured a conversational AI assistant similar to ChatGPT and the ability to insert AI generated text directly into the document. Refer to Figure \ref{fig:baseline_interface} for a screenshot. 

\section{Scenario Descriptions and Prompts}
\label{scenario_appendix}
The two task scenarios were described to the participants as follows:
\begin{itemize}
    \item \textbf{LinkedIn Post:} Imagine you're crafting a LinkedIn post to secure a copywriting job. Copywriters produce captivating, clear-cut text tailored for various advertising mediums like websites and print ads. You want to convince your network to point to relevant opportunities and form new connections.  
    \item \textbf{Email to a Professor:} Imagine you're writing an email to introduce yourself to Professor Bardley, with whom you've never communicated before. Aiming to leave a positive first impression, you're exploring multiple ways to best introduce yourself. 
\end{itemize}

For each scenario, participants generated an initial draft of roughly the same length using the following prompts:
\begin{itemize}
    \item \textbf{LinkedIn Prompt:} Help me write a LinkedIn post to find a job as a copywriter. I have some experience writing posts for a university club to ensure members stay engaged. I also took a course on copywriting last fall and want to highlight that. I am excited about writing and want to convince my connections to direct me to roles that might be a good fit or introduce me to people. Keep it within three paragraphs.
    \item \textbf{Email Prompt:} Compose an email to Professor Bardley. I've never had the opportunity to meet them, but I'm eager to make a favorable first impression. I'm enrolled in their Computational Social Science course for the upcoming fall and aspire to join their lab as a research assistant next summer. I want to convey my familiarity with their significant work on detecting misinformation on social media and developing tools to counteract it. Keep it within three paragraphs.
\end{itemize}
\end{document}